\documentclass[a4paper,showkeys,floatfix,aps,pre,longbibliography,superscriptaddress,preprint]{revtex4-1}
\usepackage{graphics,graphicx}
\usepackage{amsmath,amssymb}
\usepackage{graphics,graphicx}
\usepackage{dcolumn,bm}
\usepackage{psfrag}
\usepackage{xstring}
\usepackage{color}
\newcommand{\srm}
{\affiliation{Department of Physics, SRM University - AP,
 Andhra Pradesh - 522240, India}}

\begin{document}

\title{Critical scaling through Gini index}

\author{Soumyaditya Das}

\email{soumyaditya\_das@srmap.edu.in}
\srm
\author{Soumyajyoti Biswas}

\email{soumyajyoti.b@srmap.edu.in}
\srm

\begin{abstract}

In the systems showing critical behavior, various response functions have a singularity at the critical point. Therefore, as the driving field is tuned towards its critical value, the response functions change drastically, typically diverging with universal critical exponents. In this work, we quantify the inequality of response functions with measures traditionally used in economics, namely by constucting a Lorenz curve and calculating the corresponding Gini index. The scaling of such a response function, when written in terms of the Gini index, shows singularity at a point that is at least as universal as the corresponding critical exponent. The critical scaling, therefore, becomes a single parameter fit, which is a considerable simplification from the usual form where the critical point and critical exponents are independent. We also show that another measure of inequality, the Kolkata index, crosses the Gini index at a point just prior to the critical point. Therefore, monitoring these two inequality indices for a system where the critical point is not known, can produce a precursory signal for the imminent criticality. This could be useful in many systems, including that in condensed matter, bio- and geophysics to atmospheric physics. The generality and numerical validity of the calculations are shown with the Monte Carlo simulations of the two dimensional Ising model, site percolation on square lattice and the fiber bundle model of fracture.  
\end{abstract}
 
%\pacs{89.75.Da, 89.65.-s, 64.60.De, 75.78.Fg}

\maketitle

%\section{Introduction}
%%%%%%%%%%%%%%%%%%%%%%%%%%%%%%%%%%%%%%%%%%%%%%%%%%%%%%%%%%%%%%%%%%%%%%%%%%%%%%%%%%%%%%%%

Critical phenomena are observed in an expansive variety of physical systems undergoing equilibrium (fluids, binary mixtures, magnetic systems, superfluidity, superconductivity etc.), as well as non-equilibrium phase transitions (fracture, active particles etc.) \cite{skma}. When a system approaches a critical point by tuning a driving field $F$ towards its critical value $F_c$, a suitably defined response function $ {M}$ (e.g., derivatives of free energy) would show a singular variation of the form $ {M}\propto |F-F_c|^{-n}$. While from the universality hypothesis the value of the critical exponent $n$ remains the same within a class of systems, the critical point $F_c$ very much depends on the details of each system, thereby posing one of the major difficulties in estimating the critical exponent values \cite{crpt_book}. 

In this work, we present a framework where the critical behavior can be formulated using a measure called the Gini index ($g$), that quantifies how unequal the response of a system is near the critical point \cite{gini}. The Gini index have been used for over a century in quantifying economic inequality. However, for the critical scaling of physical quantities, the Gini index of a response function shows a singularity at a point which is at least as universal as the corresponding critical exponent. Hence the critical scaling for any unknown function becomes a one-parameter fit. We also formulate a precursory signal for an imminent critical point using a different measure of inequality, the Kolkata index ($k$) \cite{kolkata}.

Due to the singular form of $M$, small changes in $F$ can result in changes by very unequal amounts in ${M}$, depending upon the proximity to the critical point $F_c$ (here we take $F_c>0$, without loss of generality). Such highly unequal responses are ubiquitously manifested in various physical systems. For example, the drastic changes in the magnetic susceptibility near a ferro-paramagnetic transition \cite{crpt_book}, growing avalanche sizes in a stressed quasi-brittle material driven towards the failure point \cite{wiley_book}, widely varying energy releases in earthquake events due to slowly moving tectonic plates \cite{rmp}, occurrences of catastrophic desertification due to small changes in endogenous \cite{pnas} pressure, are a few instances of such unequal responses of measurable quantities near the corresponding critical points.

Given the often consequential nature of such transitions, along with estimating the critical exponent values, it is also of wide interests to predict the proximity to an imminent critical transition point \cite{ew1,ew2,ew3}. As the critical point is a non-universal quantity, one often has to resort to multi-parameter fitting, Binder cumulant calculations, machine learning based regressions or other system-specific methods in order to estimate the critical exponent values as well as the proximity to an imminent drastic change in the system \textit{i.e.,} the critical point \cite{method1,method2,method3,method4,method5,method6,method7,method8}.

Here we show that any response function $M\propto |F-F_c|^{-n}$ can be written in terms of the corresponding Gini index as $ {M}\propto |g-g_f|^{-n^*}$, where  $n^*$ is a function of $n$ and $g_f$ is either a function of $n$ or 1. We also show, through another measure of the inequality in values of $ {M}$, the so called Kolkata index ($k$), that the condition $g=k$ is satisfied for $F=F^*<F_c$ if $n>1$, with the crossing point value of the two indices approaching $1/2$ from above as $n\to\infty$, thereby the condition acts as a precursor to the approaching criticality. For the ranges of the value of $n$ that usually appears in physical systems, this crossing point values is close to $0.87$ (very weakly dependent on $n$).

%%%%%%%%%%%%%%%%%%%%%%%%%%%%%%%%%%%%%%%%%%%%%%%%%%%%%%%%%%%%%%%%%%%%%%%%%%%%%%%%%%%%%%%%%%
\begin{figure}
\includegraphics[width=9cm]{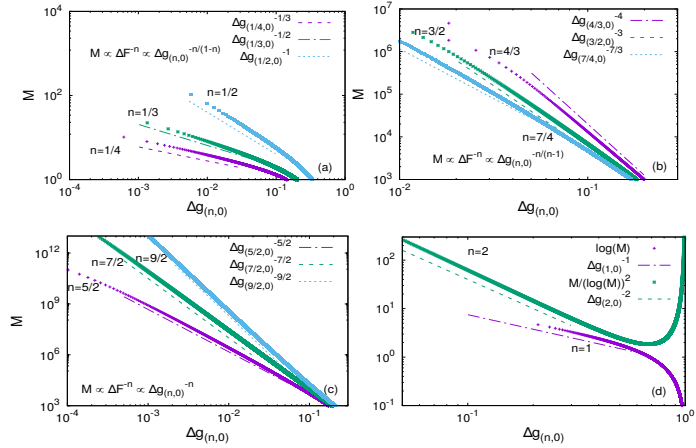}
\caption{The scaling behavior of a response function in terms of the Gini index is shown. As $M\propto \Delta F^{-n}$, the divergence with respect to the Gini index is $M\propto \Delta g_{(n,0)}^{-n^*}$, where the divergence exponent $n^*=n/(1-n)$ for $0<n<1$ ((a) showing some typical examples), $n^*=n/(n-1)$ for $1<n<2$ ((b) showing some typical examples), $n^*=n$ for $n>2$ ((c) showing some typical examples) and (d) shows the particular cases of $n=1$ and $n=2$.}
\label{fig_2}
\end{figure}
%%%%%%%%%%%%%%%%%%%%%%%%%%%%%%%%%%%%%%%%%%%%%%%%%%%%%%%%%%%%%%%%%%%%%%%%%%%%%%%%%%%%%%%%
The inequality indices $g$ and $k$ (and similar other indices) are defined using the so called Lorenz function. Lorenz function was introduced in 1905 primarily to quantify wealth inequality in an economy \cite{lorenz}. Traditionally the function $\mathcal{L}(p)$ is defined as the fraction of the total wealth of a society possessed by the poorest $p$ fraction of the population. In the present context, for a monotonically diverging response function, the function $\mathcal{L}(p)$ can be computed within an arbitrary range from $F=A$ to $F=B$ as
\begin{equation}
 \mathcal{L}(p,n,A,B) = \frac{ \int_{A}^{{A} +p(B-A)} {M}\, dF \ }{ \int_{A}^{{B}} {M}\,dF\ },
 \label{lorenz_fn}
\end{equation}
where $A<B<F_c$. Experimentally/numerically, in the ferromagnets/ Ising model for example, for a series of values of temperature below the critical point, one could compute the Lorenz function with the (unequal) values of susceptibility ($M=\chi, F=T$ there) using the above equation. For the other side of criticality, the limits need to be appropriately reversed.  By definition $\mathcal{L}(p=0)=0$ and $\mathcal{L}(p=1)=1$. Within the range $0<p<1$, $\mathcal{L}(p)$ is continuous, monotonically growing and with a positive curvature, if any. In the extreme limit if $ {M}$ is independent of $F$ i.e., the responses are always equal no matter the driving field then $\mathcal{L}(p)=p$, which is called the equality line. The departure of $\mathcal{L}(p)$ from this equality line, therefore, is a measure of the inequality in $ {M}$.

%within an interval between $aF_c$ and $bF_c$ can be written as 
%\begin{equation}
% L(p,n,a,b) = \frac{ \int_{a F_c}^{{a F_c} +p(b-a)F_c} {M}(F,F_c)\, dF \ }{ \int_{a F_c}^{{b F_c}} {M}(F,F_c)\,dF\ },
% \label{lorenz_fn}
%\end{equation}
%where $0<a < b \le 1$ with $b=1$ at the critical point. 

To put a value to this inequality \textit{i.e.}, to define an inequality index/coefficient, one needs to look at what is called the summary statistics of $\mathcal{L}(p,n,A,B)$ i.e., the $p$ dependence needs to be removed. This can be done by either integrating $\mathcal{L}(p,n,A,B)$ over the full range of $p$, or by evaluating it at a particular value of $p$. The Gini index, defined as 
\begin{equation}
   g(n,A,B)=1-2\int\limits_0^1\mathcal{L}(p,n,A,B)dp 
\end{equation}
 is an exercise of the former, while the Kolkata index, defined as the fixed point $1-k=\mathcal{L}(k,n,A,B)$, is that of the latter. The interpretation for $g$ is that it is the area between the equality line and the Lorenz curve divided by the area under the equality line (necessarily $1/2$). Therefore, it varies between $g=0$ (complete equality) to $g=1$ (just one value is non-zero). The $k$ index has the interpretation that $1-k$ fraction of the largest values account for the $k$ fraction of the total value. It is a generalization of the Pareto's law \cite{pareto}. 

We will first look at the properties of the Lorenz function and particularly the Gini index, when measured near the critical point of a system. To quantify proximity to the critical point, let us write $A=aF_c$ and $B=bF_c$. Then
from Eq. (\ref{lorenz_fn}), using the power-law variation of $ {M}$, we get ($n\ne 1$, $n\ne 2$)
\begin{equation}
    \mathcal{L}(p,n,a,b)=\frac{(1-a)^{1-n}-(1-a-p(b-a))^{1-n}}{(1-a)^{1-n}-(1-b)^{1-n}},
    \label{lorenz}
\end{equation}
 It is then straightforward to evaluate the Gini index
 \begin{widetext}
\begin{equation}
    g(n,a,b)=1-\frac{2}{(1-a)^{1-n}-(1-b)^{1-n}} \left[(1-a)^{1-n}+\frac{(1-b)^{2-n}-(1-a)^{2-n}}{(2-n)(b-a)}\right]
    %g(n,a,b)=1-2\left[ \frac{(1-a)^{1-n}}{(1-a)^{1-n}-(1-b)^{1-n}}+
    %\frac{(b-a)^{1-n}}{(2-n)\left((1-a)^{1-n}-(1-b)^{1-n}\right)}
    %\left\{ \left(\frac{1-b}{b-a}\right)^{2-n}-\left(\frac{1-a}{b-a}\right)^{2-n}  \right\}
    %\right]
    \label{gini}
\end{equation}
\end{widetext}
while the Kolkata index needs to be numerically evaluated from
\begin{equation}
    1-k(n,a,b)=\frac{(1-a)^{1-n}-(1-a-k(n,a,b)(b-a))^{1-n}}{(1-a)^{1-n}-(1-b)^{1-n}}.
    \label{kolkata}
\end{equation}
We will use the notation $g(b=1,n)=g_f$ and $k(b=1,n)=k_f$ and keep $n>0$. In the following we consider the cases $n<1$, $n>1$ and $n>2$ separately, with the corresponding consequences in the scaling form of $ {M}$. Note that the Gini index can be calculated from the $q$-th order derivative of $ {M}$ w.r.t. $F$, which is also a diverging function at $F_c$ with a different exponent. Let us, therefore, fix the notation that $\Delta g_{(\phi,q)}$ denotes the critical interval, when $g$ (and $g_f$) are calculated using the $q$-th order derivative of some response function having the original (i.e., in terms of $|F-F_c|$) exponent $\phi$. So, for example, $\Delta g_{(\alpha,1)}$ would mean that it is calculated for the first derivative of specific heat and so on. The same is true for the rescaled exponents, i.e., $\gamma_{(\beta,1)}$ would mean the susceptibility exponent appearing in the power of $\Delta g_{(\beta,1)}$, where $g$ is calculated using the first derivative of the order parameter (a diverging quantity at $F_c$). 

\noindent {\bf Case I ($0<n<1$):} Clearly, $g(b=1,n)=g_f=n/(2-n)$, which is independent of $a$. It also follows from Eq. (\ref{gini}) that for $b\to 1$, keeping upto the leading order term in $(1-b)^{1-n}$, we have  (the details of the calculations are given in the Supplemental Materials (SM))
\begin{equation}
    g(n,a,b)\approx \frac{n}{2-n}-2(1-b)^{1-n}(1-a)^{n-1},
\end{equation}
which means $|g-g_f|=\Delta g_{(n,0)} \propto (1-b)^{1-n}(1-a)^{n-1}$. Since $1-b\propto F_c-F$ and $a$ is constant,  
\begin{equation}
     {M}\propto \Delta g_{(n,0)}^{-n/(1-n)}, 
    \label{scaling:case1}
\end{equation}
with $g_f=n/(2-n)$ and $0<n<1$. See Fig. \ref{fig_2}(a) for comparisons with numerical evaluations for some typical values of $n$.

%It is interesting to note that $\beta_{(\beta,1)}=1$, i.e., the order parameter varies linearly with $\Delta g_{(\beta,1)}$ for any system near any critical point (see SM Fig. 3 for a demonstration of it in the mean-field Ising model and the FBM).

\noindent {\bf Case II ($1<n<2$):} In the limit $b\to 1$, Eq. (\ref{gini}) gives $g_f=1$. It also follows from Eq. (\ref{gini}) that upto the leading order $|g-g_f|=\Delta g_{(n,0)}  \propto (1-a)^{1-n}(1-b)^{n-1}$, which as before leads to
\begin{equation}
     {M} \propto \Delta g_{(n,0)}^{-n/(n-1)},
    \label{scaling:case2}
\end{equation}
with $g_f=1$ and $1<n<2$. See Fig. \ref{fig_2}(b) for comparisons with numerical evaluations for some typical values of $n$. See also Fig. 4(a) of the Appendix C for the manifestation of this scaling in two dimensional Ising model.

\noindent {\bf Case III ($n>2$):} Here also $g_f=1$. Then upto the leading order, $|g-g_f|=\Delta g_{(n,0)} \propto (1-b)/(1-a)$. This implies
\begin{equation}
     {M} \propto \Delta g_{(n,0)}^{-n},
    \label{scaling:case3}
\end{equation}
with $g_f=1$ and $n>2$. See Fig. \ref{fig_2}(c) for comparisons with numerical evaluations for some typical values of $n$. See also Fig. 4(b) of the Appendix C for the manifestation of this scaling in the site percolation. 

            \noindent {\bf Case IV ($n=1$ and $n=2$):} 
For $n=1$, upto the leading order $g\approx 1+\frac{2}{b}(\frac{1}{ln(1-b)})$ and with $g_f\to 1$ for $n\to 1$, $ln(M)\sim \Delta g_{n,0}^{-1}$. Similarly, for $n=2$, $g\approx 1+\frac{2}{b}((1-b)ln(1-b))$. With $g_f\to 1$ for $n\to 2$, we have $M/(ln(M))^2\sim \Delta g_{(n,0)}^{-2}$. Numerically these are verified in Fig. \ref{fig_2}(d).

%%%%%%%%%%%%%%%%%%%%%%%%%%%%%%%%%%%%%%%%%%%%%%%%%%%%%%%%%%%%%%%%%%%%%%%%%%%%%%%%%%%%%%%%%
%%%%%%%%%%%%%%%%%%%%%%%%%%%%%%%%%%%%%%%%%%%%%%%%%%%%%%%%%%%%%%%%%%%%%%%%%%%%%%%%%%%%%%%%%
\begin{figure*}
\includegraphics[width=17cm]{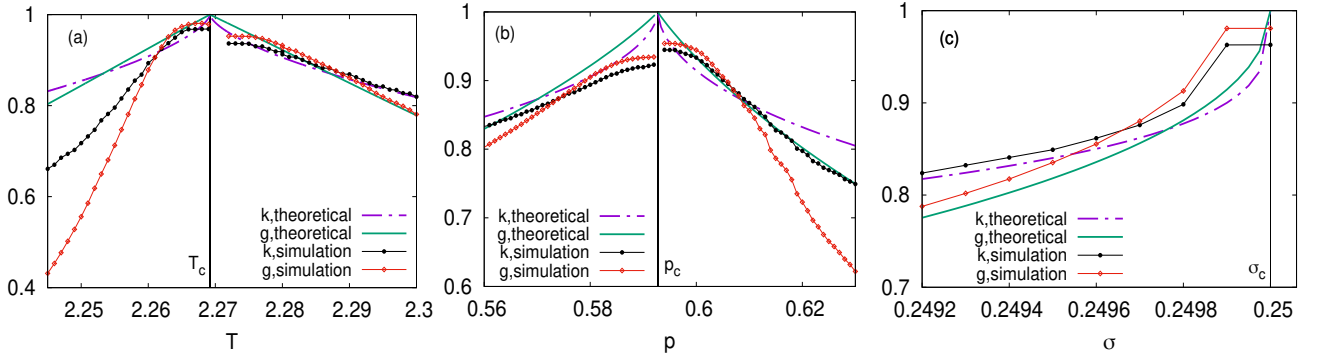}
\caption{The precursory signals from the crossing points of Gini ($g$) and Kolkata ($k$) indices. (a) The values of $g$ and $k$ are
measured for $\chi^2$ in the two dimensional Ising model from either side of the critical point (by increasing and decreasing temperature
from below and above the critical point respectively). The crossing happens close to the critical point. $g$ and $k$ do not real 1 due to finite size effect.
(b) Here the same is done for the second moment of cluster sizes for the site percolation in two dimensions. (c) Here the cube of the avalanche sizes are taken for 
the fiber bundle model ($S^3 \propto (\sigma_c-\sigma)^{-3/2}$). The crossing can only be shown here in the pre-critical regime, since there is no stable
configuration of the model for $\sigma>\sigma_c$, the catastrophic failure point. In all cases the analytical estimates are also shown, which do not match very well
since in the simulations the power-law variation is only valid very close to the critical point. But the crossing point values for $g$ and $k$ are almost independent of the associated exponent value.}
\label{fig_3}
\end{figure*}
%%%%%%%%%%%%%%%%%%%%%%%%%%%%%%%%%%%%%%%%%%%%%%%%%%%%%%%%%%%%%%%%%%%%%%%%%%%%%%%%%%%%%%%%

%%%%%%%%%%%%%%%%%%%%%%%%%%%%%%%%%%%%%%%%%%%%%%%%%%%%%%%%%%%%%%%%%%%%%%%%%%%%%%%%%%%%%%%%%
%\begin{figure*}
%\includegraphics[width=15cm]{fig3_combo_new.eps}
%\caption{The precursory signals from the crossing points of Gini ($g$) and Kolkata ($k$) indices. (a) The inequality indices $g$ and $k$ are calculated from the cube avalanche sizes $S^3(\sigma)$ of the random fiber bundle model with system size $L=100000$ with different values of lower cut-offs ($a$) in the applied load. The applied load is critical at $b=1$. Since $S(\sigma)\propto (\sigma_c-\sigma)^{-1/2}$, $S^3(\sigma)$ diverges with an exponent $3/2$, resulting in a crossing of $g$ and $k$ at a value $\sigma_e<\sigma_c$. The zoom of the crossing points are shown in (b) for $a=0.8$ {\it i.e.,} lower cut-off at 80\% of the critical point, in (c) for $a=0.84$, in (d) for $a=0.88$ and in (e) for $a=0.92$. The lines in all cases denote the corresponding analytical estimates.}
%\label{fig_3}
%\end{figure*}
%%%%%%%%%%%%%%%%%%%%%%%%%%%%%%%%%%%%%%%%%%%%%%%%%%%%%%%%%%%%%%%%%%%%%%%%%%%%%%%%%%%%%%%%

In the above cases, we have written a generic response function near any critical point in such a way that the critical exponent ($n^*$) and the critical point ($g_f$) are equally universal. We have calculated $g$ from one side of the critical point for the diverging response function, but it is extendable to the other side of the critical point (see Fig. 3 of Appendix A). Also, it follows that the corresponding coefficients of such a diverging function are expected to be the same on both sides. 

Note that, for practical purposes, when the proximity to the critical point is {\it a-priori} not known, for a series of values of the driving parameter $F$, one can calculate a series of values of $g$ (from Eq. (\ref{gini})) for a response function. An estimate of the critical point could be made beforehand by noting the maximum of $g$ (see Appendix C, specifically Fig. 7). These $g$ values can then  be fitted using Eq. (\ref{scaling:case1}), (\ref{scaling:case2}) or (\ref{scaling:case3}), which is then a single parameter fit, since $g_f$ is solely dependent on $n$, the exponent value. This is a considerable simplification from the usual situation where the critical point ($F_c$) and the critical exponent ($n$) are independent.   

It is useful to revisit the implications of the critical scaling using $g$ on (a) the finite size scaling, to show what is expected for a simulation study with finite system sizes, and (b) precursor to critical point, for a practical application of the inequality measures in a variety of systems.

\noindent {\bf (a) Finite size scaling:} A characteristic feature of second order phase transition is the divergence of a correlation length $\xi$ at the critical point $F_c(\infty)$ (in the infinite system size limit), where $F_c$ can be critical temperature ($T_c$) in Ising model, percolation threshold ($p_c$) in percolation or critical applied stress ($\sigma_c$) in the fiber bundle model (FBM) of fracture etc. In a finite systems, however, near $F_c(L)$ the role of $\xi$ is taken over by the linear system size $L$:
\begin{equation}
    |F-F_c(\infty)| \propto \xi^{-\frac{1}{\nu}} \xrightarrow{} |F_c(L)-F_c(\infty)| \propto L^{-\frac{1}{\nu}}.
\end{equation}
Recalling that $|g-g_f|\propto c(n)|F-F_c|^{\theta}$, where $c(n)$ is only dependent on $n$ and $\theta=1-n$ for $0<n<1$, $\theta=n-1$ for $1<n<2$ and $\theta=1$ for $n>2$, 
at the critical point of a finite system (of linear size $L$) we would have
\begin{equation}
|g_f(L)-g_f(\infty)| \propto L^{-\frac{\theta}{\nu}},
\end{equation}
with the values of $\theta$ depending on $n$ as mentioned above. This is numerically verified for the Ising model on square lattice (see Fig. 8).

\noindent {\bf (b) Precursor to critical point:} The closed form of the Kolkata index $k$ is not possible for arbitrary $n$ (see Eq. (\ref{kolkata})). However, its numerical evaluation shows the remarkable property that for $n>1$, $k$ becomes equal to $g$ at two points: one is the trivial point where $g_f=k_f=1$ at the critical point, but the other point (say, $g^*=k^*$ at $F=F^*$) is necessarily below the critical point and usually very close to it (see Fig. 9 in Appendix E). Therefore, for any system approaching a critical point (from either side, if possible), monitoring $g$ and $k$ for a sufficiently strongly diverging response function ($n>1$) would indicate an imminent critical point when the two quantities become equal and has a value smaller than 1 (see Fig. \ref{fig_3}). 

Having a reliable precursory signal to an imminent critical point is a crucial issue in many physical systems including fracture, environmental catastrophe, market crash etc.   In the case of fracture \cite{wiley_book}, this issue have been addressed in several different ways, including using inequality indices \cite{pre1,pre2,front1}. 

Here we take three paradigmatic examples, the Ising model and the site percolation problem on square lattices and the fiber bundle model of fracture and show that $g^*=k^*$ at $F=F^*<F_c$ (where $F=T$ in Ising model, $F=p$ in percolation and $F=\sigma$ in FBM) is a reliable precursor to critical point on both sides of criticality for the first two models and for one side in the case of FBM (since there is no stable state on the other side of criticality in this case). 

First we consider the dynamics of the fiber bundle model for fracture, which is viewed as a critical phenomena for several decades.
 It is a threshold activated cellular automata type model that reproduces many features of fracture dynamics (see \cite{fbm_rmp} for a review), including the intermittent scale-free avalanche dynamics in disordered quasi-brittle materials.  With $N$ elements (fibers) carrying a load $W$, 
the mean field version of the model is analytically tractable. For a mild restriction on the failure threshold (load beyond which a fiber breaks and redistributes its load to the remaining fibers) probability distributions of the individual fibers, the fraction of surviving fibers $U(\sigma)=N(\sigma)/N$ for an applied load per fiber $\sigma=W/N$ has the form $U(\sigma)=U(\sigma_c)+D(\sigma_c-\sigma)^{1/2}$, where $D$ is a constant that depends on the distribution function and $\sigma_c$ is the critical load beyond which the system collapses \cite{moreno}. One can then consider the response function 
\begin{equation}
S(\sigma)=\left|\frac{dU}{d\sigma}\right|\propto (\sigma_c-\sigma)^{-1/2}, 
\end{equation}
which has the physical interpretation of the avalanche size (if a constant amount of load $d\sigma$ is added to the system every time it comes to a stable state).
% then 
%the Lorenz function in Eq. (\ref{lorenz}) becomes $L(p,a,b,n=1/2)=\frac{\sqrt{1-a}-\sqrt{1-a-p(b-a)}}{\sqrt{1-a}-\sqrt{1-b}}.$ 
For detecting the precursory signal from the Gini and Kolkata indices (i.e. to make them cross), we need a function that diverges with an exponent higher than 1. We consider the function $S^3(\sigma)$, which will diverge with an exponent $3/2$. A higher power would still work, but will give a precursory signal earlier, eventually leading to the trivial limit where precursor is set as soon at $F>0$ (see Fig. 9 in Appendix E). We numerically evaluate $S^3(\sigma)$ from the simulation data. Then we calculate the inequality indices $g_{(3/2,0)}(a,b,n=3/2)$ and $k_{(3/2,0)}(a,b,n=3/2)$ and found that they cross at a point prior to the critical point (see Fig. \ref{fig_3} (c)). The crossing point, therefore can serve as an indicator to imminent critical point (catastrophic breakdown in this case) irrespective of the threshold distribution function. 

%The closer the measurements of response (susceptibility/avalanche) is made to the (tuned) critical point, more it resembles the response near the self-organized critical (SOC) point.

Note that in an SOC state, the system is always very close to the critical point. Its response statistics are generally scale-free. 
 It is analytically known for the FBM that the avalanche size distribution exponent value is the same for both the (mean field) SOC case \cite{njp} and for the avalanches occurring only very close to the (tuned) critical point \cite{hemmer}. Note that the crossing point value $g^*=k^*\approx 0.87$, which is almost independent of the divergence exponent, is what was numerically observed in simulations \cite{manna} of SOC models (including FBM) and the real data of many systems assumed to be in the SOC state \cite{evol,front1}. This near-universal observation can now be argued from the above to be a consequence of the measurements of inequality indices (from Eq. (\ref{gini}) and (\ref{kolkata})) of the corresponding response functions very close to the critical point. 
%It is also worth noting that the method assumes a power-law scaling of the response functions, which is usually valid very close to the critical point. Also, presence of confluent singularities might hamper detecting $g_f$.  

The generality of this precursory signal can be seen by applying it for the two dimensional Ising model and site percolation on square lattice. For the Ising model, the susceptibility ($\chi$) diverges with an exponent $\gamma=7/4$ \cite{skma}. While $g$ and $k$ are expected to cross for this, the crossing point is expected to be very close to the critical point (see Fig. 9 in Appendix E). So, we take $\chi^2$ instead (diverging with an exponent $7/2$), for which the crossing points for $g$ and $k$ can be seen (Fig. \ref{fig_3} (a)) from both sides of the critical point. Similarly, for the site percolation on square lattice, the second moment of the cluster size distribution diverges with an exponent $43/18$ \cite{stf}. Here also, the crossing of $g$ and $k$ could be seen prior to the critical point on both sides of the critical point (Fig. \ref{fig_3} (b)).

In conclusion, inequality measures of diverging response functions near a critical point enable a super-universal representation of such functions (see Eqs. \ref{scaling:case1}, \ref{scaling:case2}, \ref{scaling:case3}) that are free from the non-universal, model specific critical point.  It also allows for a precursory signal of an approaching criticality, which is crucial in many systems. The analytical results are verified through numerical simulations of the two dimensional Ising model, site percolation on square lattice and the fiber bundle model of fracture, but these are applicable to any equilibrium or non-equilibrium critical phenomenon.

\appendix

\section{Gini index near the critical point ($b\to 1$) for different values of $n$}

\begin{figure}[h]
\includegraphics[width=10cm]{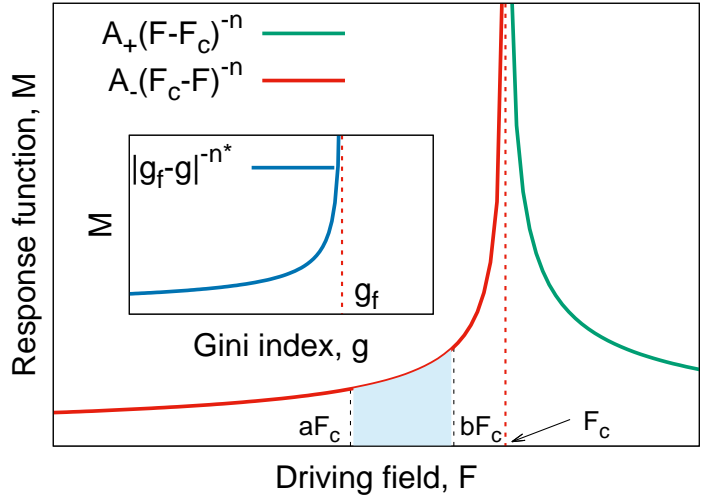}-
\caption{A Schematic diagram showing the singular behavior of a response function ($M$) in terms of the driving force, $F$ (main figure) and the Gini index, $g$ (inset). The response function diverges at $F=F_c$ with a critical exponent $n$. The inset shows the divergence of the same function with respect to the Gini index calculated within the interval $aF_c$ to $bF_c$, with $g(b=1)=g_f$ as the point of divergence, which is either only a function of $n$ (for $n<1$) or 1 (for $n\ge 1$). The divergence exponent ($n^*$) is given by $n^*=n/(1-n)$ for $0<n<1$, $n^*=n/(n-1)$ for $1<n<2$, $n^*=n$ for $n>2$ and for $n=1$ and $n=2$ logarithmic corrections are seen (see Fig. 2 in the main text). The Gini index will always be less than 1, hence the divergence on both sides with respect to $F$ can be mapped to that with respect to $g$ on one side ($g<g_f\le 1$).}
\label{schem}
\end{figure}

As mentioned in the main text, we are interested in calculating the inequality of a diverging response function $M \propto |F-F_c|^{-n}$ by varying $F$ from $aF_c$ to $bF_c$, where $a< b\le 1$. The Gini index calculated for $b=1$, the critical point, is called $g_f$ (see Fig. \ref{schem}), which is independent of $a$. 

We have the general expression (Eq. (4) in the main text) for the Gini index as

\begin{widetext}
\begin{equation}
     g(n,a,b)=1-\frac{2}{(1-a)^{1-n}-(1-b)^{1-n}} \left[(1-a)^{1-n}+\frac{(1-b)^{2-n}-(1-a)^{2-n}}{(2-n)(b-a)}\right]
    \label{gini}
    \end{equation}
\end{widetext}

On simplification which gives,

\begin{equation}
    g(n,a,b)=1-\frac{2}{\left\{ 1-\left(\frac{1-b}{1-a}\right)^{1-n}\right\}}\left[ 1+\frac{1-a}{(2-n)(b-a)}\left\{ \left(\frac{1-b}{1-a}\right)^{2-n} -1\right\}\right]
    \label{gini_sim}
\end{equation}

%%%%%%%%%%%%%%%%%%%%%%%%%%%%%%%%%%%%%%%%%%%%%%%%%%%%%%%%%%%%%%%%%%%%%%%%%%%%%%%%%%%%%%%%
%%%%%%%%%%%%%%%%%%%%%%%%%%%%%%%%%%%%%%%%%%%%%%%%%%%%%%%%%%%%%%%%%%%%%%%%%%%%%%%%%%%%%%%%

\noindent {\bf Case I ($0<n<1$):} As  $\left(\frac{1-a}{b-a}\right)  \approx 1 $, the Eq. (\ref{gini_sim}) becomes,
 \begin{widetext}
     \begin{align}
         g(n,a,b)\approx 1-\frac{2}{\left\{ 1-\left(\frac{1-b}{1-a}\right)^{1-n}\right\}}\left[ 1-\frac{1}{2-n}\left\{1- \left(\frac{1-b}{1-a}\right)^{2-n}\right\}\right]
     \end{align}
\end{widetext}

Putting $x=\frac{1-b}{1-a}$, and in the limit of small $x$

    \begin{eqnarray}
         g(n,a,b) &\approx& 1-\frac{2}{\left( 1-x^{1-n}\right)}\left[ 1-\frac{1}{2-n}\left(1- x^{2-n}\right)\right] \nonumber \\
        &\approx& 1-2\left[(1-x^{1-n})^{-1} - \frac{1}{2-n}\left\{\frac{1- x^{2-n}}{1-x^{1-n}}\right\}\right] \nonumber \\
        & \approx& 1-2(1+x^{1-n})\left[1-\frac{1}{2-n}\right] \nonumber \\
        &\approx& \frac{n}{2-n} - \frac{2-2n}{2-n}\left(\frac{1-b}{1-a}\right)^{1-n}. 
    \end{eqnarray}
    Note that in the expansion above, we have kept upto the first term, which is true only when $n$ is not very close to 1. The $n=1$ case is treated separately below, also for $n$ close to either side of 1, there needs to be correction terms in the scaling.
    With $g_f=\frac{n}{2-n}$, the above equation gives
    \begin{equation}
        \Delta g =|g-g_f| \propto  \left(\frac{1-b}{1-a}\right)^{1-n}.
    \end{equation}

\noindent {\bf Case II ($1<n<2$):} 
As  $\left(\frac{1-a}{b-a}\right)  \approx 1$, $\left\{ 1-\left(\frac{1-b}{1-a}\right)^{1-n}\right\}\approx - \left(\frac{1-b}{1-a}\right)^{1-n}$ and $\left\{1- \left(\frac{1-b}{1-a}\right)^{2-n}\right\} \approx 1 $ [Note that this approximation is not valid when n is very close to 2] , thus Eq. (\ref{gini_sim}) becomes,

\begin{equation}
    g(n,a,b) \approx 1+2\left(\frac{1-b}{1-a}\right)^{n-1}\left[1-\frac{1}{2-n}\right].
\end{equation}

With $g_f=1$,
\begin{equation}
     \Delta g= |g-g_f| \propto  \left(\frac{1-b}{1-a}\right)^{n-1}
\end{equation}

\noindent {\bf Case III ($n>2$):} Here $\left(\frac{1-a}{b-a}\right) \approx 1$ , $\left\{ 1-\left(\frac{1-b}{1-a}\right)^{1-n}\right\}\approx - \left(\frac{1-b}{1-a}\right)^{1-n}$ and $\left\{1- \left(\frac{1-b}{1-a}\right)^{2-n}\right\} \approx - \left(\frac{1-b}{1-a}\right)^{2-n} $,thus Eq. (\ref{gini_sim}) becomes,

\begin{eqnarray}
g(n,a,b) &\approx& 1+2\left(\frac{1-b}{1-a}\right)^{n-1}\left[1-\frac{1}{2-n}\left(\frac{1-b}{1-a}\right)^{2-n}\right] \nonumber \\
    &\approx& 1-\frac{2}{2-n}\left(\frac{1-b}{1-a}\right). 
\end{eqnarray}
With $g_f=1$,   

\begin{equation}
     \Delta g= |g-g_f| \propto  \left(\frac{1-b}{1-a}\right)
\end{equation}

\noindent {\bf Case IV ($n=1$ and $n=2$):} 
\noindent{\bf For $n=1$:} The Lorenz function can be written as 
\begin{equation}
    \mathcal{L}(a,b,p)=\frac{\ln\left(\frac{1-a-p(b-a)}{1-a}\right)}{\ln\left(\frac{1-b}{1-a}\right)}.
\end{equation}
Then the Gini index can be calculated as
\begin{equation}
    g(a,b,n=1)=1+\frac{2(1-a)}{(b-a)\ln\left(\frac{1-b}{1-a}\right)}\left[\left(\frac{1-b}{1-a}\right)\ln\left(\frac{1-b}{1-a}\right)-\left(\frac{1-b}{1-a}\right)+1\right],
\end{equation}

which can then be approximated near the critical point as,
\begin{equation}
    g(a,b,n=1)\approx 1+\frac{2}{\ln\left(\frac{1-b}{1-a}\right)}.
\end{equation}

With $g_f=1$,
\begin{equation}
    \Delta g =|g_f-g|\propto - \left[\ln\left(\frac{1-b}{1-a}\right)\right]^{-1}
\end{equation}
which means, 
\begin{equation}
    \ln M \propto \Delta g ^{-1}
\end{equation}

 \noindent{\bf For $n=2$:}                    

\begin{eqnarray}
    \lim_{n\to 2}g(n,a,b) &=& 1+2\left(\frac{1-b}{1-a}\right)\left[ 1+\lim_{n\to 2}\left\{\frac {\left(\frac{1-b}{1-a}\right)^{2-n} -1}{2-n}\right\}\right] \nonumber \\
    &=& 1+2\left(\frac{1-b}{1-a}\right)\left[1+\ln\left(\frac{1-b}{1-a}\right)\right].
\end{eqnarray}
Again with $g_f=1$,
\begin{equation}
     \Delta g \propto\ \left(\frac{1-b}{1-a}\right)\ln\left(\frac{1-b}{1-a}\right).
\end{equation}
Hence, 
\begin{equation}
    \frac{M}{(\ln M)^{2}} \propto (\Delta g)^{-2}
\end{equation}

\section{\bf The Rushbrooke inequality under the Gini index scaling}
Since any response function can be rewritten in terms of the Gini index, it is useful to see what happens to the scaling relations that exists between the original exponents. 

 If we have a set of exponents $0<\alpha<1$, $0<\beta<1$ and $1<\gamma<2$, say for the three dimensional Ising model, they will satisfy the Rushbrooke equality $\alpha+2\beta+\gamma=2$. Suppose we consider the specific heat, which has the exponent $\alpha$ for its divergence. One can calculate $g$ from the diverging susceptibility and write $(g_f-g) \propto |T-T_c|^{1-\alpha}$ (see Eq. (7) in the main text and the discussions above that equation). Therefore, $|T-T_c| \propto (g_f-g)^{1/(1-\alpha)}$. Here of course, according to our notation, $(g_f-g)=\Delta g_{(\alpha,0)}$, the first value of the subscript indicating that the divergence exponent was $\alpha$ and the second value (0) indicating that we did not take any derivative of the response function (here specific heat) before calculating $g$. Once this is achieved, it is possible to write any other response function in terms of $\Delta g_{(\alpha,0)}$, and the associated exponent will just be the original exponent divided by $(1-\alpha)$. For example, susceptibility will be $\chi \propto |T-T_c|^{-\gamma}\propto \Delta g_{(\alpha,0)}^{-\gamma/(1-\alpha)}$. Similarly, the order parameter (magnetization) $m \propto \Delta g_{(\alpha,0)}^{\beta/(1-\alpha)}$ and the specific heat $C \propto \Delta g_{(\alpha,0)}^{-\alpha/(1-\alpha)}$.
So we have $\gamma_{(\alpha,0)}=\gamma/(1-\alpha)$, $\beta_{(\alpha,0)}=\beta/(1-\alpha)$ and $\alpha_{(\alpha,0)}=\alpha/(1-\alpha)$.

In a similar way, one could have started from the Gini index of diverging susceptibility. But since $1<\gamma<2$, the scaling mentioned in Eq. (8) in the main text would be followed, and the rescaled exponents could be found from dividing the original exponents by $\gamma-1$, so for example $C \propto \Delta g_{(\gamma,0)}^{-\alpha/(\gamma-1)}$ and so on. So we have $\gamma_{(\gamma,0)}=\gamma/(\gamma-1)$, $\beta_{(\gamma,0)}=\beta/(\gamma-1)$ and $\alpha_{(\gamma,0)}=\alpha/(\gamma-1)$.

For the order parameter, however, one has to take a derivative (w.r.t $T$) first, before calculating $g$. Given that $0<\beta<1$, $dm/dT$ will diverge with an exponent less that 1, implying the applicability of the scaling in Eq. (7) in the main text. The rescaled exponents can readily be written down, for example $\chi \propto |T-T_c|^{-\gamma} \propto \Delta g_{(\beta,1)}^{-\gamma/\beta}$. A few things to note here. First, we needed a derivative, hence the second index is 1. Secondly, the divergence exponent of $dm/dT$ is $1-\beta$, hence the rescaling is by $1/(1-(1-\beta))=1/\beta$. So we have $\gamma_{(\beta,1)}=\gamma/\beta$, $\beta_{(\beta,1)}=1$ and $\alpha_{(\beta,1)}=\alpha/\beta$.

Now, if  $\alpha+2\beta+\gamma=2$, then it follows that 
\begin{eqnarray}
\alpha_{(\alpha,0)}+2\beta_{(\alpha,0)}+\gamma_{(\alpha,0)}=\frac{2}{1-\alpha} &>&2 \nonumber \\
\alpha_{(\gamma,0)}+2\beta_{(\gamma,0)}+\gamma_{(\gamma,0)}=\frac{2}{\gamma-1} &>& 2 \nonumber \\
\alpha_{(\beta,1)}+2\beta_{(\beta,1)}+\gamma_{(\beta,1)}=\frac{2}{\beta} &>& 2.
\end{eqnarray}
Generally, the Rushbrooke inequality
\begin{equation}
    \alpha_{(\phi,q)}+2\beta_{(\phi,q)}+\gamma_{(\phi,q)}\ge 2,
\end{equation}
 where $\phi \in (\alpha,\beta,\gamma)$ and $q\ge 0$, is valid, since for $q>1$ there will not be any change in the exponents (see Eq. (9) in the main text).

\section{\bf Numerical verification of scaling through Gini index in the Ising model, site percolation and fiber bundle model}
Here we show the numerical corroborations of the scalings shown using the critical Gini index interval in the main text calculations.

\subsection{Scaling of susceptibility in Ising model and cluster size in percolation}
\begin{figure}
\includegraphics[width=16cm]{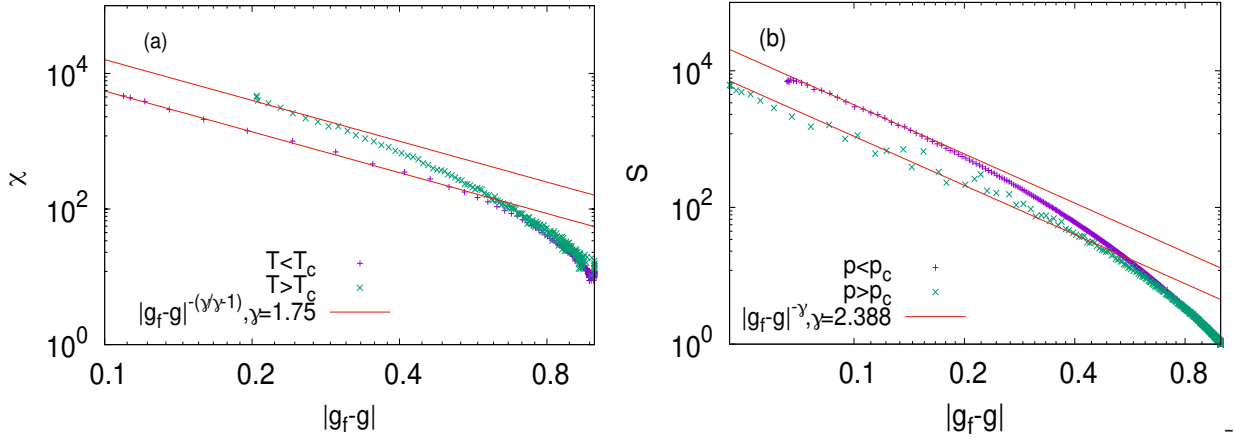}-
\caption{The scaling of (a) the susceptibility for the two dimensional Ising model ($\chi \sim |T-T_c|^{-7/4}$), simulated for $L=700$ and (b) the second moment of the cluster size distribution for 
the site percolation on the square lattice ($S\sim |p-p_c|^{-43/18}$), simulated for $L=500$, are shown for the critical Gini index intervals. For the Ising model, the scaling predicted in Eq. (8)
in the main text is seen on both sides of the critical point and for the site percolation, the scaling predicted in Eq. (9) in the main text is seen on both sides of the critical point.}
\label{ising_perco}
\end{figure}

For the two dimensional Ising model, the susceptibility scales as $\chi \sim |T-T_c|^{-\gamma}$ with $\gamma=7/4$ on both sides of the critical point. Therefore, 
when written in terms of the critical Gini index interval, it should follow the scaling predicted in Eq. (8) in the main text. This implies a scaling of the form 
$\chi \sim |g-g_f|^{-\gamma/(\gamma-1)} \sim |g-g_f|^{-2.333}$. In Fig. \ref{ising_perco}(a) we show that indeed such a scaling is observed on both sides of the critical point. The range of the scaling is rather small, 
which is due to the limited range of $g$ and the fact that due to finite size scaling (discussed later) the critical Gini index interval does not become zero. 

Similarly, for the second moment of the cluster size distribution of site percolation on a square lattice, there is a strong divergence near the critical point
$S \propto \sum\limits_s s^2n_s \sim |p-p_c|^{-43/18}$. So, it is expected to follow the scaling reported in Eq. (9) in the main text. Therefore, we should have
$S \sim |p-p_c|^{-43/18}$, which is what is seen in Fig. \ref{ising_perco}(b) on both sides of the critical point. 
%%%%%%%%%%%%%%%%%%%%%%%%%%%%%%%%%%%%%%%%%%%%%%%%%%%%%%%%%%%%%%%%%%%%%%%%%%%%%%%%%%%%%%%%
%%%%%%%%%%%%%%%%%%%%%%%%%%%%%%%%%%%%%%%%%%%%%%%%%%%%%%%%%%%%%%%%%%%%%%%%%%%%%%%%%%%%%%%%

%%%%%%%%%%%%%%%%%%%%%%%%%%%%%%%%%%%%%%%%%%%%%%%%%%%%%%%%%%%%%%%%%%%%%%%%%%%%%%%%%%%%%%%%
\begin{figure}
\includegraphics[width=8cm]{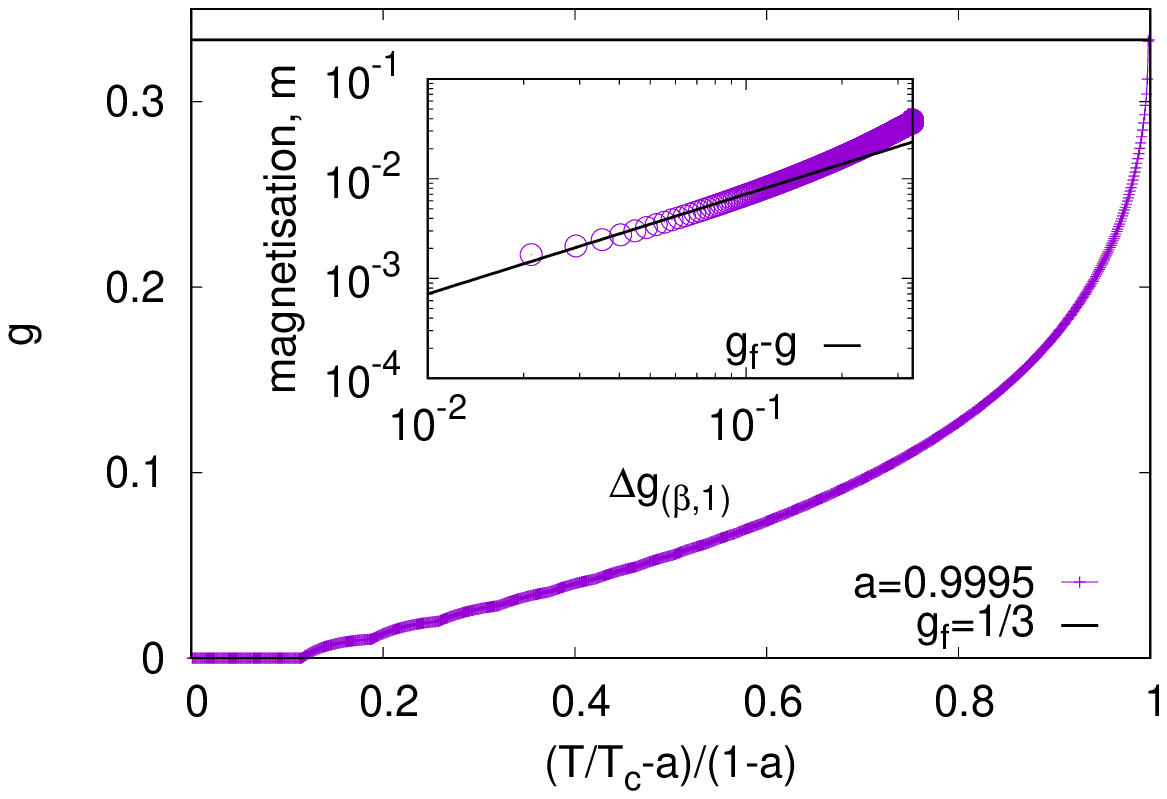}-
\includegraphics[width=8cm]{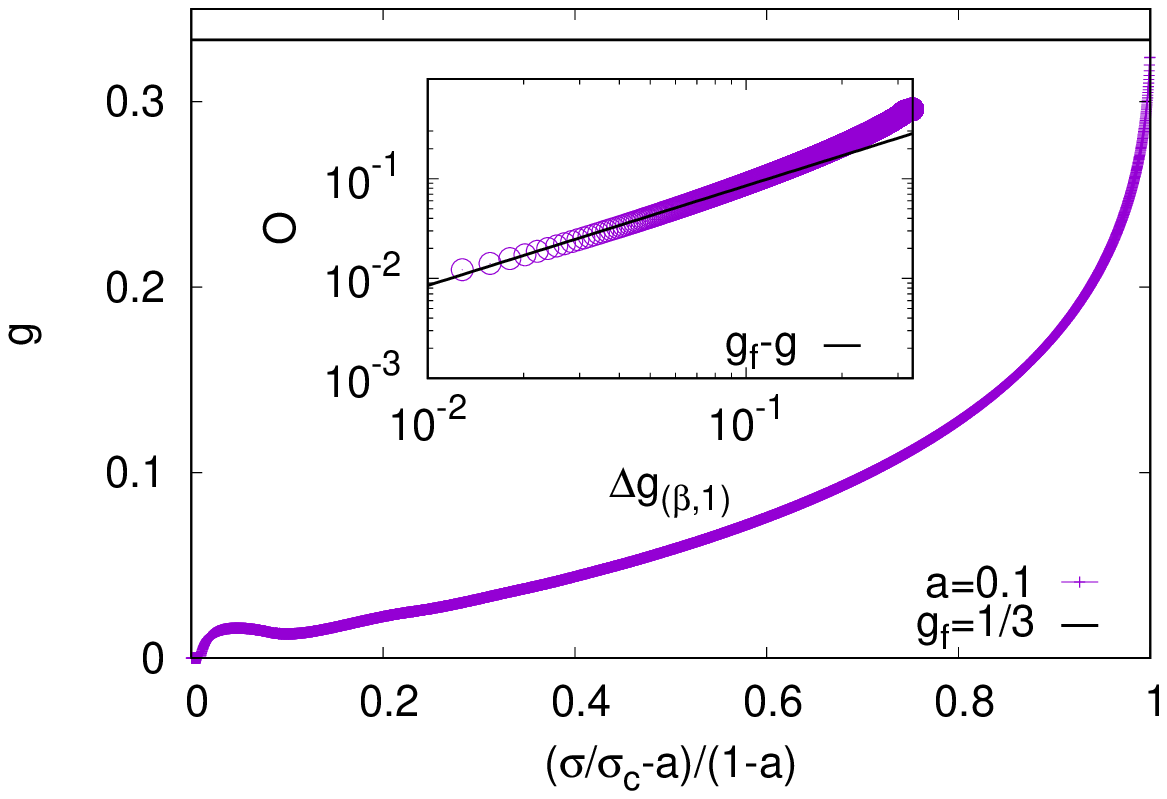}
\caption{The scaling of the mean-field Ising model and the Fiber Bundle Model (FBM) order parameters using the corresponding Gini indices. (a) The main plot shows the evolution of the Gini index with the driving field $T$ (adjusted for the cut-off $a$). The inset shows the linear scaling with $\Delta g_{\beta,1}$ as discussed in the text. (b) The main plot shows the evolution of the Gini index with the driving field $\sigma$ (adjusted for the cut-off $a$). The inset shows the linear scaling with $\Delta g_{\beta,1}$ as discussed in the text. }
\label{ising_fig}
\end{figure}
%%%%%%%%%%%%%%%%%%%%%%%%%%%%%%%%%%%%%%%%%%%%%%%%%%%%%%%%%%%%%%%%%%%%%%%%%%%%%%%%%%%%%%%%

\subsection{Scaling of the order parameter: Mean field Ising and FBM cases}
In the mean-field approximation, the magnetization shows singular behavior near the critical point with the critical exponent ($\beta$) $1/2$ i.e, $m \sim |T-T_c|^{1/2}$ for $ (T<T_c) $. Now, a `response function' ($M$) can be constructed by taking the derivative of $ m $ w.r.t T which also shows singular behavior in the form $M = |\frac{dm}{dT}| \sim |T-T_c|^{-n}$ with $n=\frac{1}{2}$. To calculate the Lorenz function ($L$) and eventually the Gini index ($g$), the same procedure can be followed which was discussed in the main text. Hence, $\Delta g_{(\beta,1)} =|g-g_f| \propto  \left(\frac{1-b}{1-a}\right)^{1/2}$ i.e, $\Delta g_{(\beta,1)} \propto |T-T_c|^{1/2}$ with $g_f=1/3$ implies a linear relationship between $m$ and $\Delta g_{(\beta,1)}$ independent of $n$ (except of course, $g_f$ will depend on $n$ but the linearity is nevertheless maintained) i.e, $m \propto \Delta g_{(\beta,1)}$, as can be seen in Fig. \ref{ising_fig}.

%%%%%%%%%%%%%%%%%%%%%%%%%%%%%%%%%%%%%%%%%%%%%%%%%%%%%%%%%%%%%%%%%%%%%%%%%%%%%%%%%%%%%%%%
\begin{figure}
\includegraphics[width=16cm]{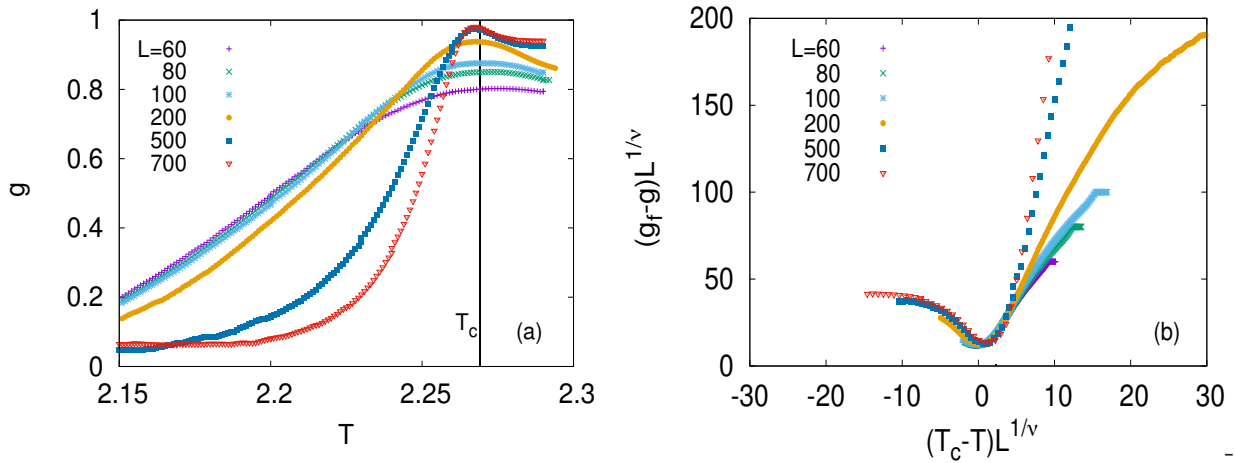}-
\caption{The plot of $g$ (calculated for $\chi^2$ on the two dimensional Ising model) versus temperature show a maximum close to the
critical point (a). The finite size scaling is shown in (b), where the collapse is expected very close to 0 in the x-axis.}
\label{fss_gini_merged}
\end{figure}
%%%%%%%%%%%%%%%%%%%%%%%%%%%%%%%%%%%%%%%%%%%%%%%%%%%%%%%%%%%%%%%%%%%%%%%%%%%%%%%%%%%%%%%%

%%%%%%%%%%%%%%%%%%%%%%%%%%%%%%%%%%%%%%%%%%%%%%%%%%%%%%%%%%%%%%%%%%%%%%%%%%%%%%%%%%%%%%%%
\begin{figure}
\includegraphics[width=16cm]{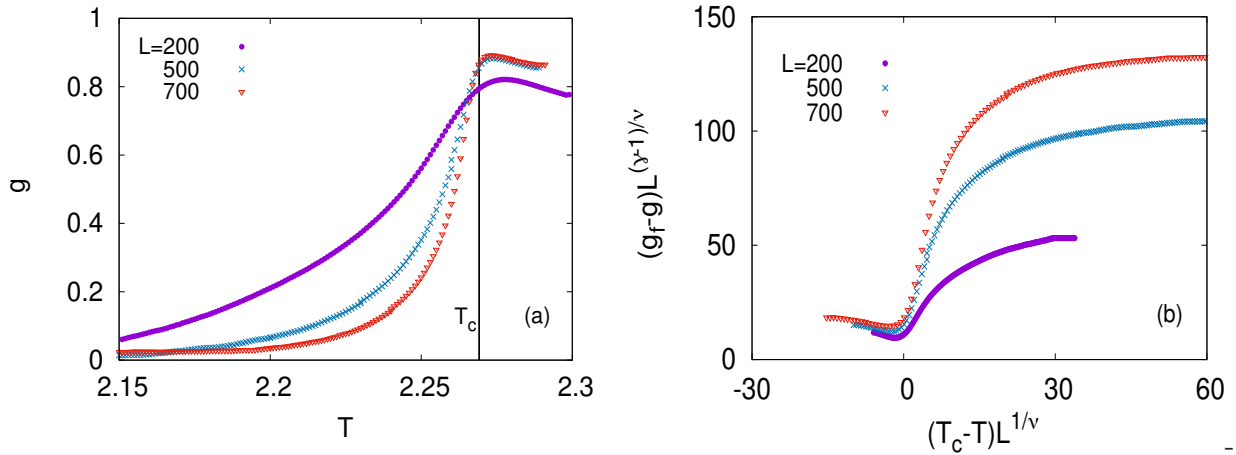}-
\caption{(a) The plot of $g$ (calculated for $\chi$ on the two dimensional Ising model) with $T$. (b) The finite size scaling of $g$, calculated for $\chi$ on the two dimensional Ising model is shown. The scaling works better for the higher system sizes, as
opposed to a much better fit near 0 seen for $\chi^2$ in Fig. \ref{fss_gini_merged}.}
\label{fss_gini_2}
\end{figure}
%%%%%%%%%%%%%%%%%%%%%%%%%%%%%%%%%%%%%%%%%%%%%%%%%%%%%%%%%%%%%%%%%%%%%%%%%%%%%%%%%%%%%%%%

The same can be done for the Fiber Bundle Model, for which the order parameter, for a broad class of threshold distributions, behaves as $O=U^*(\sigma_c)-U*(\sigma)=D(\sigma_c-\sigma)^{1/2}$ \cite{fbm_rmp}, where $D$ is a constant that depends on the threshold distribution of the fibers. The above mentioned linear scaling, i.e., $O \propto \Delta g_{(\beta,1)}$ is still valid, which can be seen from the simulation of the fiber bundle model (in Fig. \ref{ising_fig}) with a threshold distribution that is uniform in $(0,1)$. 

Note here that the power law scaling of the order parameter for the FBM is valid in the entire range of the driving field, whereas for the Ising model the power law scaling is only near the critical point. Therefore, a lower cutoff is needed for the Ising model that is close to 1, whereas for the FBM it can even be zero (here we take $a=0.1$). Also, the definition of the order parameter in the FBM needs the quantity $U^*(\sigma_c)$, which is the stationary value of the fraction of surviving fiber just at the critical load. It can be argued then the knowledge of the critical point is inherent in the definition of the order parameter and cannot just be eliminated by the framework here. However, there is at least one alternative definition of the order parameter for the FBM, with the exact same scaling behavior, that does not require the knowledge of $U^*(\sigma_c)$, called the branching ratio \cite{branch}. Therefore, all of these could be done in terms of the branching ratio, which avoids the complication mentioned above.  

%%%%%%%%%%%%%%%%%%%%%%%%%%%%%%%%%%%%%%%%%%%%%%%%%%%%%%%%%%%%%%%%%%%%%%%%%%%%%%%%%%%%%%%%
\begin{figure}
\includegraphics[width=8cm]{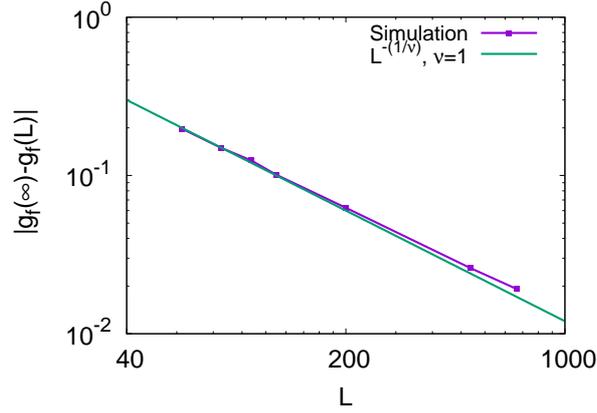}-
\caption{The finite size scaling relation obtained in Eq. (11) in the main text is verified for the
Gini index calculated for $\chi^2$ on two dimensional Ising model.}
\label{fss_gini}
\end{figure}
%%%%%%%%%%%%%%%%%%%%%%%%%%%%%%%%%%%%%%%%%%%%%%%%%%%%%%%%%%%%%%%%%%%%%%%%%%%%%%%%%%%%%%%%

\section{Finite size scaling}
The finite size scaling was discussed in the main text resulting in the form of Eq. (11) there. 
It, therefore, implies that $|g-g_f|L^{\theta/\nu}$ versus $|F-F_c|L^{1/\nu}$ plot will collapse at
the critical point showing a minimum. For the same reason, just measuring $g$ as a function of the driving parameter will give a maximum
at the critical point (in the infinite system size limit). We show that scaling for $\chi^2$ of the two
dimensional Ising model in Fig. \ref{fss_gini_merged}. The scaling of $\chi$ is also shown in Fig. \ref{fss_gini_2}, which is not as good as that seen for 
$\chi^2$.

This scaling could be used for determining the critical point in a system.

Also, the expected scaling in Eq. (11) of the main text is numerically verified for $\chi^2$ (with $\theta=1$ here) in Fig. \ref{fss_gini}. 

%%%%%%%%%%%%%%%%%%%%%%%%%%%%%%%%%%%%%%%%%%%%%%%%%%%%%%%%%%%%%%%%%%%%%%%%%%%%%%%%%%%%%%%%
\begin{figure}
\includegraphics[width=8cm]{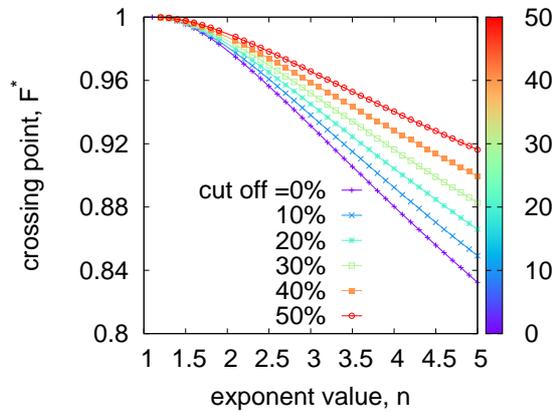}-
\caption{The crossing of $g$ and $k$ happens for $F=F^*$. Here we show the variation of $F^*$ ($F_c=1$) with the divergence exponent $n$, for different distances (lower cut-offs) for which the $g$ and $k$ values are calculated.}
\label{f_star}
\end{figure}
%%%%%%%%%%%%%%%%%%%%%%%%%%%%%%%%%%%%%%%%%%%%%%%%%%%%%%%%%%%%%%%%%%%%%%%%%%%%%%%%%%%%%%%%

\section{Signals of critical point through Gini index}
As mentioned in the main text, one clear precursory signal for the imminent critical point is the crossing of $g$ and $k$ prior to the critical point, when measured for a sufficiently strongly diverging response function, say susceptibility or its higher powers. The proximity of the crossing point value of the driving field $F^*$ to the
critical value of the field $F_c$ can be quantified as a function of the divergence exponent $n$ ($=\gamma$ for $\chi$, $=2\gamma$ for $\chi^2$), and for different values of the
distance ($a$) from which $g$ is measured (recall that $g$ is measured for a segment $F=aF_c$ to $F=bF_c$, with $a<b$ and $g=g_f$ for b=1, where $g$ becomes independent of $a$).

In Fig. \ref{f_star} the variation of $F^*$ is shown with $n$ for different values of $a$. We have kept $F^*=1$. So, when $a$ is close to 1 (of course $b$ is even closer), the crossing point $F^*$ increases (tending towards 1). Also, as was reported in the main text, the crossing point value $g^*=k^*\approx 0.87$ for the values of $n$ that we encounter. This supports the observation that for a wide class of SOC systems, $g$ and $k$ cross near the critical point, since in SOC we are always close to the critical point.  

Finally, as shown for the finite size scaling, just calculating $g$ from one side of the critical point (then crossing over to the other side by passing over some maximum value of the response function) shows a peak close to the critical point. The finite size scaling of that, as mentioned above, is a good determination of the critical point for any system. 

\begin{acknowledgments}
 The authors thank Bikas K. Chakrabarti and Parongama Sen for discussions at various stages of this work and comments on the manuscript. The simulations were performed on the HPCC Surya cluster at SRM University - AP.
\end{acknowledgments}


\begin{thebibliography}{99}
\bibitem{moreno}
One could also use something called a branching ratio as the order parameter, as defined in \cite{branch}.
\end{thebibliography}
 \end{document}